\documentclass[twocolumn,preprintnumbers,amsmath,amssymb,aps,prb]{revtex4}
\usepackage{graphicx}
\begin{document}

\title{Clogging and Depinning of Ballistic Active Matter Systems in Disordered Media}
 
\author{C. Reichhardt and C.J.O. Reichhardt}
\affiliation{Theoretical Division and Center for Nonlinear Studies,
Los Alamos National Laboratory, Los Alamos, New Mexico 87545, USA}

\date{\today}

\begin{abstract}
We numerically examine ballistic active disks driven through a random obstacle array.  Formation of a pinned or clogged state occurs at much lower obstacle densities for the active disks than for passive disks.  As a function of obstacle density we identify several distinct phases including a depinned fluctuating cluster state, a pinned single cluster or jammed state, a pinned multicluster state, a pinned gel state, and a pinned disordered state.  At lower active disk densities, a drifting uniform liquid forms in the absence of obstacles, but when even a small number of obstacles are introduced, the disks organize into a pinned phase-separated cluster state in which clusters nucleate around the obstacles, similar to a wetting phenomenon.  We examine how the depinning threshold changes as a function of disk or obstacle density, and find a crossover from a collectively pinned cluster state to a disordered plastic depinning transition as a function of increasing obstacle density.  We compare this to the behavior of nonballistic active particles and show that as we vary the activity from completely passive to completely ballistic, a clogged phase-separated state appears in both the active and passive limits, while for intermediate activity, a readily flowing liquid state appears and there is an optimal activity level that maximizes the flux through the sample.
\end{abstract}

\maketitle

\vskip 2pc

\section{Introduction}
There is a wide range 
of soft and hard matter systems that can be
modeled  as collectively interacting
particles which, when driven over quenched disorder,
exhibit pinning-depinning behavior as well as
transitions between different types   
of sliding regimes \cite{1,2}. Such dynamics occur for vortex 
motion in type-II superconductors \cite{3,4}, sliding charge density waves \cite{5}, 
depinning of classical Wigner
crystals \cite{6,7}, current driven motion of skyrmions in chiral
magnets \cite{8,9}, colloids interacting with random \cite{10,11,12,13,14,15} or 
ordered substrates \cite{16}, sliding in frictional systems \cite{17}, 
magnetic domain wall motion \cite{18,19}, erosion \cite{20}, granular matter \cite{21,22}, 
driven pattern forming systems
\cite{23}, geophysical models of plate tectonics \cite{24}, 
and the motion of dislocations in crystalline materials \cite{N}.
The substrate may be
random, ordered, or partially ordered, and it can be 
modeled as localized pinning sites
with a finite trapping strength
or as 
impenetrable obstacles.
Under an applied drive,
the particles exhibit a variety of pinned and moving
order-disorder
transitions
that can  be characterized 
by the moving structure, pattern formation
features, changes in the velocity force curves, and
fluctuation phenomena \cite{1,2}.
In the systems listed above,
the particles themselves
are passive or experience only thermal fluctuations,
so the driving is strictly externally applied;
however, recently a growing number of studies
have focused on what are called active matter systems
containing self-driven particles with an activity that is often
modeled as arising from driven diffusive or run-and-tumble dynamics \cite{25,26}.
In the absence of a substrate,
active disks exhibit  a
transition from a uniform gas or liquid state 
to a phase-separated or
cluster state consisting of a high density 
solid coexisting with a low density active gas \cite{27,28,29,30,31,31a,32}. 
This transition occurs for fixed activity
as a function of increasing density or for fixed density with increasing
activity level.
Active matter systems 
have also been been 
studied in the context of particle shape effects \cite{26,33}, active rotators \cite{34}, 
passive and active mixtures \cite{35,36},
boundary effects \cite{37,38,39,40,41}, and ratchet effects \cite{42,43,44}.

Several studies have examined different types of active matter systems coupled with 
ordered \cite{45,46,47} or 
disordered substrates \cite{31a,48,49,50,51,52,53,54,55,56,57,58,59,60,61,62,63}. 
Numerical simulations show
that when
a drift force is applied to
run-and-tumble disks moving through a random obstacle array,
the flux through the system 
is non-monotonic as a function of activity level,
indicating that
there is an optimal
run length or run correlation time
that maximizes the flux of disks through the obstacles \cite{31a}.
At small run lengths, the
disks behave thermally
and easily become trapped behind the obstacles, giving a low disk flux.
As the activity level or run length increases, the disks
can more readily move around the obstacles, increasing the flux;
however, when the run length is too large,
a self-pinning effect occurs in which the disks self-cluster around the obstacles,
reducing the flux
\cite{31a,62}.
As a result,
the flux is maximized when the disks are active and the self-clustering is weak,
while it is reduced 
when the activity becomes high
enough for significant clustering or self-trapping to occur.
When the run lengths are very long, the average flux under an applied drive
is strongly reduced, but it never reaches zero
since there are still long-time dynamical rearrangements
that produce
avalanches or intermittent flow of disks in the direction of the drift force \cite{62}. 
Analytic and theoretical
studies of active systems without a 
drift force also indicate that there is an optimal activity level
that maximizes the diffusion in the system \cite{61}.    
If the obstacles are replaced by pinning sites,
the onset of clustering can have the opposite effect of
increasing rather than decreasing the flux, since the
clusters act like large rigid objects that are poorly trapped
by individual pinning sites \cite{57}.
When the activity is low,
a uniform liquid state appears and individual disks
can be trapped readily by individual pinning sites \cite{57}.
These works also showed that within the phase separated state,
increasing the number of obstacles or pinning sites
produces a disorder-induced transition from a 
phase-separated state to a uniform disordered state \cite{55,57}.    

In addition to the run-and-tumble models,
several simulation studies of swarming or flocking active systems
with quenched disorder
show that there is an optimal noise level for the appearance of flocking
\cite{48,51},
and that increasing the disorder strength can induce a transition from
a flocking to a non-flocking state \cite{50}.
Experiments on colloidal rollers that 
exhibit flocking behavior have also
revealed a transition from a drifting 
flocking state to a non-flocking state as a function 
of increasing obstacle density, where the flow becomes increasingly filamentary 
as the disordered state is approached \cite{54}.   

In these systems, the activity is stochastic
in nature.
For run-and-tumble particles,
after each run interval the particle randomly reorients and runs in a new direction,
while in driven diffusive models, there is a noise term controlling the
rate of rotational diffusion.
The deterministic limit of active matter is
purely ballistic flow where the swimming direction of each
particle is permanently fixed.  This can be achieved
in run-and-tumble disk systems by setting the running time to infinity,
or in driven diffusive systems by setting the rotational diffusion
coefficient to zero \cite{49}.
In the deterministic regime, the particles can form
a cluster state even at very low densities,
and Bruss {\it et al.} \cite{64} argue that phase separation occurs
when the mean time between collisions is smaller than the mean
duration of an individual collision.
Previous studies \cite{49} of an active ballistic system  
showed that the disks can form a frozen cluster state
where almost all the fluctuations are lost. This frozen state 
arises due to the lack of stochastic behavior in the system, and the
resulting cluster 
can be regarded as an absorbed state.
In contrast, 
at long but finite run times, the cluster
can gradually evolve over time due to the possibility of rare stochastic
reorganizations \cite{62}.

Previous studies of the active ballistic limit did not include an
applied drift force, so no
pinning-depinning phenomena occurred.
In this work we consider active ballistic
disks
driven through an obstacle array,
and we measure the average 
drift mobility $\langle V\rangle$ of the disks
in the direction of the driving force $F_{D}$.
We find that the system can evolve toward
a completely pinned or clogged state with $\langle V\rangle \approx 0$.
We describe our simulation in Sec.~II.
In Sec.~III we compare the active pinning or clogging to
that observed in the zero activity or passive limit. 
The critical density of obstacles
needed to induce the formation of
a clogged state
is much higher in the passive system than in the active ballistic system,
and 
we show that in general the active ballistic disks are much more susceptible
to
forming clogs than the passive disks.
In the active system,
clogging is associated with the formation
of a cluster state.
As a function of increasing obstacle density, we identify four phases:
a sliding state with either a fluctuating cluster or a liquid structure;
a pinned single cluster or jammed state consisting of a
large cluster held in place by a small number of obstacles;
a pinned multicluster
state
containing several distinct pinned clusters;
and a pinned disordered state in which the disk density remains
spatially uniform.
As the active disk density increases, we also find what we term a
pinned gel state where the disks form a percolating labyrinth structure. 
For low active disk densities where a flowing uniform liquid appears
in the absence of obstacles, we find that introduction of a small number of
obstacles causes a transition to
a pinned cluster state, which we compare to the active wetting of clusters
around an obstacle.  
In Sec.~IV we examine the velocity-force relations and
the behavior of the critical depinning force $F_{c}$. 
The depinning threshold for a cluster is finite, and there is a pronounced
increase in $F_c$ at the transition from
the pinned  single cluster state to the pinned disordered state,  
reminiscent of the
``peak effect''
observed in superconducting vortex systems at
a transition from a collectively pinned ordered or quasiordered
vortex crystal to a vortex glass state \cite{2,3}.
In Sec.~V we show how passive clogging can be
connected to the ballistic clogging limit by
considering finite but 
increasing run times for run-and-tumble active disks in obstacle arrays. 
In both the passive and ballistic clogged states, the disks
form a cluster,
while between these two limits a flowing liquid structure appears.
There is an optimal run time at which the disk flux is maximized,
and the flux
decreases with increasing run length until the disks form
a completely clogged state at infinite run times.
We discuss
how these results can be related
to granular jamming transitions \cite{65,66,67} and jamming in systems
with quenched disorder \cite{68,69}, where
the activity or the obstacle density represent an additional set of 
parameters that can be used to induce a jammed state. 
In Sec.~VI we summarize our results.  

\section{Simulation}
We consider a two-dimensional system with periodic boundary conditions
in the $x$ and $y$-directions containing
$N_{a}$ active or mobile disks that interact through a stiff
repulsive harmonic spring,
${\bf F}_{s} = k(d-2R)\Theta(d - 2R){\hat {\bf d}}$, 
where $d$ is the distance between two disks, ${\bf d}$ is the displacement vector,
$k=100$ is the spring constant, and the disk radius is $R = 0.5$.
For this value of $k$ the 
disk-disk overlaps remain very small, allowing us to define the
active disk density in terms of the area covered by the disks,
$\phi_{a} = N_{a}\pi R^2/L^2$, where the system size is $L = 100$. In 
the limit of no activity and no obstacles, the
disks form a hexagonal lattice at $\phi_{a} = 0.9$.
In addition to the mobile active disks, we introduce
$N_{\rm obs}$ obstacles that are identical to the active disks but are
permanently fixed in place.
The obstacles are initially placed in a hexagonal lattice
and are randomly diluted until we reach the desired obstacle density of
$\phi_{\rm obs} = N_{\rm obs}\pi R^2/L^2$.
Placing the obstacles in an initial hexagonal lattice ensures that there
is a fixed minimum distance between any two obstacles,
thereby avoiding the
rare obstacle density fluctuations such as large gaps or tight
obstacle clustering 
that can arise from a completely random
obstacle placement
and
dominate the dynamics.
The total disk coverage of both active and fixed disks is
$\phi_{\rm tot} = \phi_{\rm obs} + \phi_{a}$.  
The dynamics of the active disks is obtained from
the following overdamped equation of motion:
\begin{equation}
\eta \frac{ d{\bf r}_i}{dt} = {\bf F}_{s,a}^i + 
{\bf F}_{m}^i + {\bf F}_{s, \rm obs}^i + {\bf F}_{D}. 
\end{equation}
Here $\eta = 1.0$ is the damping constant and the
interaction with other active disks
${\bf F}_{s,a}$ has the form of the stiff spring repulsion ${\bf F}_s$.   
Each active disk has a  motor force
${\bf F}_{m}^i$
that is constant in magnitude and oriented in a randomly chosen direction.
In run-and-tumble systems, the motor force orientation is held fixed during
the run time $\tau_{r}$, after which a new random orientation is chosen for
the next running time.
In the ballistic limit, we set $\tau_{r}$ to infinity so that the running direction
never changes.
The forces from the obstacles ${\bf F}_{s, \rm obs}$ are also given
by ${\bf F}_s$,
and the external driving force ${\bf F}_{D}=F_D{\bf \hat x}$ is applied
uniformly to all active disks.
We also consider the passive particle limit in which
${\bf F}_{m} = 0$
and the only driving force is the externally applied ${\bf F}_D$.
To initialize the system, after establishing the location of the
obstacles we place 
mobile disks with artificially reduced radii
in nonoverlapping locations 
and allow the mobile
disks to rearrange while gradually expanding the radii to the final value of
$R = 0.5$.  With this method we can reach disk densities up to
$\phi_{\rm tot} = 0.86$.    

To characterize the system we measure the average drift velocity of the disks
$\langle V\rangle =
\langle N_{a}^{-1}\sum^{N_a}_{i=0} {\bf v} _{i}\cdot {\hat {\bf x}}\rangle $
in the direction of the drive.
We wait
$1\times 10^6$ simulation time steps for the system to settle
into a steady state and then average over an additional
$9\times10^6$ simulation time steps to obtain
$\langle V\rangle$.
We have found that increasing the waiting time
produces negligible changes in the results.    
In this work, unless otherwise noted we set
$F_{D} = 0.05$, $|{\bf F}_{m}| = 0.5$, and $\tau_r=\infty$;
however, we also consider varied $F_{D}$ and finite  $\tau_{r}$.

\begin{figure}
\includegraphics[width=3.5in]{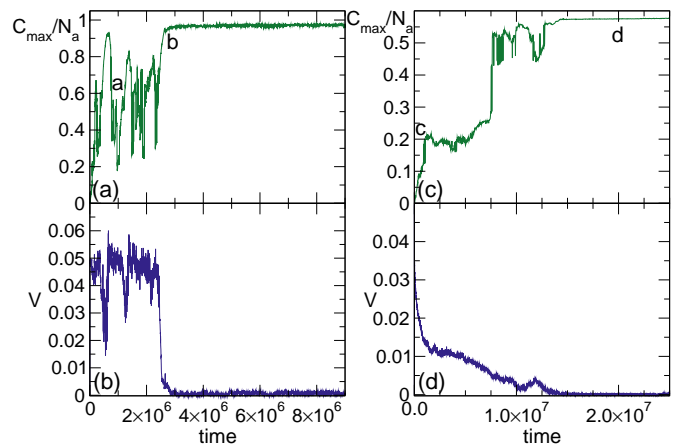}
\caption{
  (a,b) An active ballistic system with an area coverage of
  $\phi_{a}=0.3495$ active disks
  and $\phi_{\rm obs}=0.00393$ obstacles
  for a total  $\phi_{\rm tot} = 0.3534$.
  An external drift force of $F_{D} = 0.05$ is applied in  the
  positive $x$-direction.
  (a) The fraction of disks in the largest cluster, $C_{\rm max}/N_a$, versus time
  in simulation time steps.
  (b) The drift velocity $V$ of the active disks vs time.
  There is a transition from a fluctuating cluster state that is drifting in
  the direction of drive, illustrated in Fig.~\ref{fig:2}(a),
  to a pinned single cluster, shown in Fig.~\ref{fig:2}(b).
  At the transition, $C_{\rm max}/N_a$ abruptly increases to a value close to one
  and $V$ simultaneously drops nearly to zero.
  The letters {\bf a} and {\bf b} indicate the times corresponding to the images in
  Fig.~\ref{fig:2}(a,b).
(c) $C_{\rm max}/N_a$ and (d) $V$ vs time for the same
system in the passive $|{\bf F}_m|=0$ limit at $\phi_{\rm tot} = 0.3534$
and $\phi_{\rm obs} = 0.1178$.
For this value of $\phi_{\rm tot}$, the system can reach a pinned or clogged state
only when $\phi_{\rm obs} \geq 0.098$.
The system evolves over time into a pinned state, with a gradual
drop in $V$ accompanied by a gradual increase in $C_{\rm max}/N_a$.
The initial unclogged state at the time marked {\bf c} is illustrated in
Fig.~\ref{fig:2}(c),
while the $V=0$ pinned state at the time marked {\bf d} is shown in
Fig.~\ref{fig:2}(d).
} 
\label{fig:1}
\end{figure}

\begin{figure}
\includegraphics[width=3.5in]{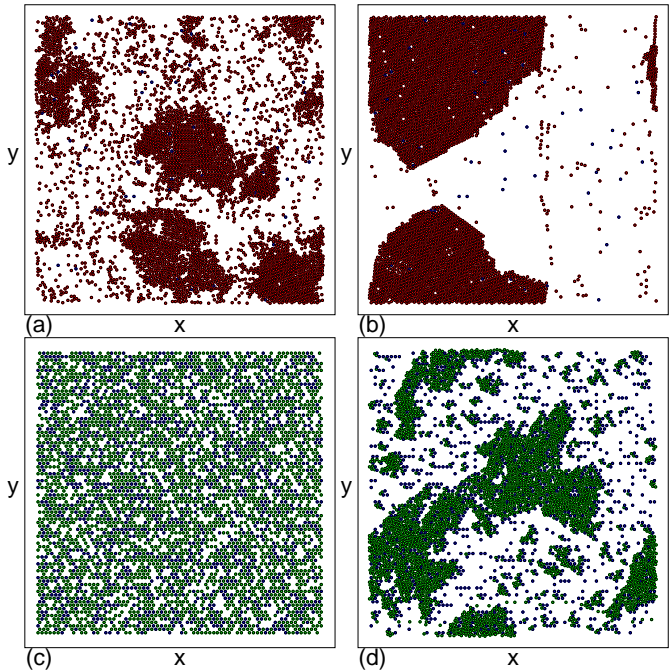}
\caption{ 
  (a,b) The active ballistic disk positions (red circles) and obstacle
  locations (blue circles) for the system in
  Fig.~\ref{fig:1}(a,b) with $\phi_{\rm tot} = 0.3534$, $\phi_{\rm obs} = 0.00393$, and
  a drift force $F_D=0.05$ applied in the positive $x$ direction.
  (a) Depinned fluctuating clusters appear at
  time $1\times10^6$, marked {\bf a} in Fig.~\ref{fig:1}(a).
  (b) The pinned single cluster state at
  time $3\times 10^6$, marked {\bf b} in Fig.~\ref{fig:1}(a).
  (c,d) Passive disk positions (green circles) and obstacle
  locations (blue circles)
  for the system from Fig.~\ref{fig:1}(c,d) with
  $\phi_{\rm tot} = 0.3534$ and $\phi_{\rm obs} = 0.1178$.
  (c) The initial flowing state at the time marked {\bf c} in Fig.~\ref{fig:1}(c).
  (d) The clogged or pinned state at  time $3\times 10^6$,
  marked {\bf d} in Fig.~\ref{fig:1}(c).
} \label{fig:2}
\end{figure}
      
\section{Phases of Active Ballistic Disks}
In Fig.~\ref{fig:1}(a) we plot the fraction
$C_{\rm max}/N_a$
of active disks in the largest cluster versus time in simulation time steps
for a system with $F_{D} = 0.05$, $N_{a} =  4450$, and $N_{\rm obs}=50$,
giving an active disk density of  $\phi_{a}  = 0.3495$,
an obstacle density of
$\phi_{\rm obs} = 0.00393$,
and an overall density of
$\phi_{\rm tot} = 0.3534$.
The largest cluster size $C_{\rm max}$ is defined as
the largest number of disks in direct contact with each 
other as determined using the cluster identification algorithm
described in Ref.~\cite{70}.
Figure~\ref{fig:1}(b) shows the corresponding
disk velocity $V=N_a^{-1}\sum_{i=0}^{N_a}{\bf v}_i \cdot {\bf \hat x}$ versus time.
Both $C_{\rm max}/N_a$
and $V$ exhibit strong fluctuations
during the first $2.5\times 10^6$ time steps, indicating that
the system
is in a dynamic unpinned fluctuating state.
The disks
then become trapped in a single pinned cluster,
as shown by the sudden jump in $C_{\rm max}/N_a$ to $C_{\rm max}/N_a \approx 1.0$
which is accompanied by a drop in $V$ to nearly zero.
There are still some small fluctuations in both $C_{\rm max}/N_a$ and $V$
due to the presence of a small number of freely
running disks that do not join the cluster.
In Fig.~\ref{fig:2}(a) we show a snapshot of the
depinned fluctuating clusters at
time $1.0\times10^6$, where
the active disks form temporary clusters.
Here $V= 0.046$, which is close to the
expected obstacle-free value of $V = 0.05$.
A pinned single cluster state appears
at time $3\times 10^6$, as illustrated
in Fig.~\ref{fig:2}(b),
where the active disks form a single large
immobile cluster.
We find that even a small number of obstacles
($N_{\rm obs}/N_{a} = 0.011$)
can produce a pinned state for active ballistic disks.
In contrast, for passive disks at the same total density of
$\phi_{\rm tot} = 0.3534$,
the system does not reach a pinned or clogged state
until $N_{\rm obs}/N_{a} \geq 0.384$,
indicating that nearly 35 times more
obstacles are required to pin the passive disks compared to the active ballistic disks.

In Fig.~\ref{fig:1}(c,d) we plot
$C_{\rm max}/N_a$ and $V$
versus time for a passive $|{\bf F}_m|=0$ system
with $\phi_{\rm tot} = 0.3534$ and
$\phi_{\rm obs} = 0.1178$, where the disks reach a
pinned or clogged state.
In contrast to the active ballistic system in
Fig.~\ref{fig:1}(a,b),
the pinned state does not appear abruptly; instead, the passive disks
continuously evolve toward the pinned state over time,
with a growing number of pinned clusters gradually
emerging as indicated by the steady
increase in $C_{\rm max}/N_a$ and the gradual decrease  in $V$.
In Fig.~\ref{fig:2}(c)
we illustrate the initial
uniform spatial distribution of the passive disks,
while in Fig.~\ref{fig:2}(d) we show a snapshot of the disk positions
in the clogged state at a time of $2\times 10^6$.
Unlike the active ballistic disks in Fig.~\ref{fig:2}(b), the passive disks
in Fig.~\ref{fig:2}(d) 
do not form a single clump
but
instead assemble into a number of smaller clumps.
This indicates  that the passive and active clogged or pinned states
are very different in nature.

\begin{figure}
  \includegraphics[width=3.5in]{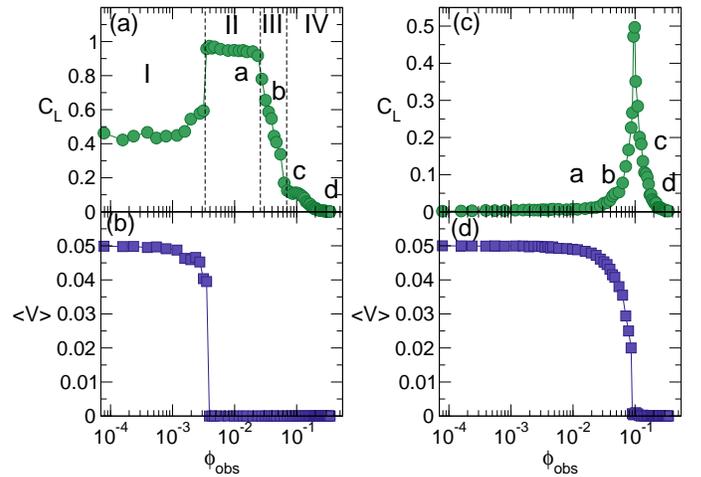}
\caption{
  (a) $C_{L}$ and (b) $\langle V\rangle$ vs
  obstacle density $\phi_{\rm obs}$
  for active ballistic disks
  with $F_{D} = 0.05$ and $\phi_{\rm tot} =  0.3534$.
  For $\phi_{\rm obs} < 0.0039$, $\langle V\rangle$ is
  finite and the system is in phase I$_{\rm fc}$, the depinned
  fluctuating cluster state, while at the transition to phase II,
  the pinned single cluster state, $\langle V\rangle$ drops to zero.
  Phase II extends from
  $0.0039 \leq \phi_{\rm obs} < 0.025$ and is illustrated
  in Fig.~\ref{fig:4}(a).
  For $0.025 \leq \phi_{\rm obs} < 0.065$,  the system is in phase III,
  a pinned multicluster state, illustrated in Fig.~\ref{fig:4}(b),
  while for $\phi_{\rm obs} \geq 0.065$ the system is in phase IV, the
  pinned disordered phase
  illustrated in Fig.~\ref{fig:4}(c,d) at $\phi_{\rm obs} = 0.0942$
  and $\phi_{\rm obs}=0.1963$.
  The labels {\bf a} to {\bf d} in (a) indicate the values of $\phi_{\rm obs}$
  at which the images in Fig.~\ref{fig:4} are obtained.
  (c) $C_{L}$ and (d) $\langle V\rangle$ vs $\phi_{\rm obs}$
  for passive disks at the same $F_D$ and $\phi_{\rm tot}$.
  Here
  there are only two phases: a plastic flow state
  $\phi_{\rm obs} < 0.098$, and a completely pinned or clogged state
  for $\phi_{\rm obs} \geq 0.098$.
  There is a peak in $C_{L}$ at $\phi_{\rm obs}=0.098$ where $\langle V\rangle$
  drops to zero.
  The labels {\bf a} to {\bf d} in (c) indicate the values of $\phi_{\rm obs}$
  at which the images in Fig.~\ref{fig:5} are obtained.
}
\label{fig:3}
\end{figure}

\begin{figure}
\includegraphics[width=3.5in]{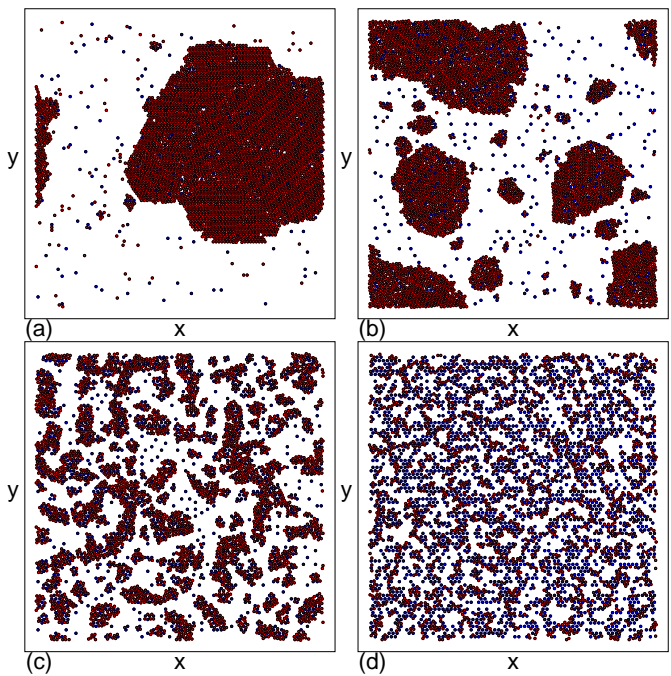}
\caption{
  The active ballistic disk positions (red circles) and obstacle
  locations (blue circles) for the
  system in Fig.~\ref{fig:3}(a,b)
  with $F_D=0.05$ and $\phi_{\rm tot}=0.3534$
  obtained at the values of $\phi_{\rm obs}$
  marked by the letters {\bf a} to {\bf d} in Fig.~\ref{fig:3}(a).
  (a) The pinned single cluster phase II at $\phi_{\rm obs} = 0.01178$.
  (b) The pinned multicluster phase III at $\phi_{\rm obs} = 0.039$.
  (c) The pinned disordered phase IV at $\phi_{\rm obs}=0.0942$
  consists of a group of small clusters.
  (d) The pinned disordered phase IV at $\phi_{\rm obs} = 0.1963$
  is composed of even smaller clusters.
}
\label{fig:4}
\end{figure}

In Fig.~\ref{fig:3}(a,b) we plot $C_{L}=\langle C_{\rm max}/N_a\rangle$ and
$\langle V\rangle$ versus $\phi_{\rm obs}$ for the active ballistic system
from Fig.~\ref{fig:1}(a) 
with $\phi_{\rm tot} = 0.3534$
at $F_{D} = 0.05$.
For $\phi_{\rm obs} < 0.0039$ the system
is in a depinned fluctuating cluster state,
labeled phase I$_{\rm fc}$, where $\langle V\rangle$ is finite
and the long-time average of $C_{\rm max}/N_a$ is $C_L\approx 0.5$.
For $0.0039 \leq \phi_{\rm obs} < 0.025$, we find a pinned single cluster state,
denoted phase II, with
$C_{L} > 0.9$ and $\langle V\rangle \approx 0.0$.
In Fig.~\ref{fig:4}(a) we illustrate phase II
at $\phi_{\rm obs} = 0.01178$.
Phase II can be viewed as a pinned jammed state
in which a single clump has nucleated around the obstacles
and acts as a rigid solid.  Active disks that are not adjacent to obstacles
are pinned or prevented from moving
by other active disks through contact interactions,
so the collective pinning of the clump
is dominated by disk-disk interactions
rather than by direct disk-obstacle interactions.
Since we are using monodisperse disks rather than the bidisperse disk mixture
commonly studied in jammed systems,
the particles
forming the cluster have
a substantial amount of hexagonal or crystalline ordering,
whereas typical 2D jammed systems form amorphous rather than
polycrystalline packings
\cite{66,68};
however, in both our system and the jamming systems,
it is the
disk-disk contact interactions
that cause the system to act like a solid that can be pinned by a small number of obstacles.
Previous work \cite{49} on active ballistic systems
revealed similar large-scale frozen cluster states, but did not include
an external drift force.
In the present study, the formation of the single frozen cluster
in the presence of a drift force results in a pinned state.

For $0.025 \leq \phi_{\rm obs} < 0.065$ in Fig.~\ref{fig:3}(a,b),
we observe a pinned multicluster state termed phase III in which
$\langle V\rangle = 0$ where $C_{L}$ decreases from
$C_L=0.9$ to $C_L=0.12$ with increasing $\phi_{\rm obs}$.
A snapshot of the disk positions in
phase III at $\phi_{\rm obs}=0.039$ appears in Fig.~\ref{fig:4}(b).
For $\phi_{\rm obs} \geq 0.065$,
the system is in phase IV, a pinned disordered state with $C_L<0.15$
in which the disks form numerous small clumps
that gradually decrease in size with increasing
$\phi_{\rm obs}$, 
as illustrated in Fig.~\ref{fig:4}(c) at $\phi_{\rm obs} = 0.0942$
and in Fig.~\ref{fig:4}(d) at $\phi_{\rm obs} = 0.1963$.

\begin{figure}
\includegraphics[width=3.5in]{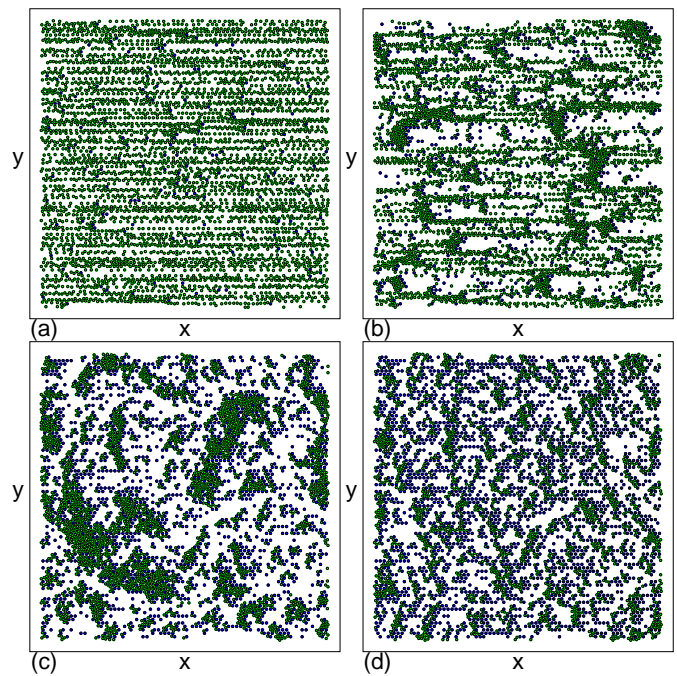}
\caption{
Passive disk positions (green circles) and obstacle locations (blue circles)
for the system in Fig.~\ref{fig:4}(c,d) with $F_D=0.05$ and $\phi_{\rm tot} = 0.3534$
obtained at the values of $\phi_{\rm obs}$ marked by the letters
{\bf a} to {\bf d} in Fig.~\ref{fig:3}(c).
(a) The flowing state at $\phi_{\rm obs} = 0.01178$.
(b) At $\phi_{\rm obs} = 0.055$, clusters begin to form.
(c) The clogged state at $\phi_{\rm obs} = 0.14137$.
(d) The clogged state at $\phi_{\rm obs} = 0.1963$.
}
\label{fig:5}
\end{figure}

We plot $C_L$ and $\langle V\rangle$ versus $\phi_{\rm obs}$ for passive
disks with $\phi_{\rm tot}=0.3534$ and $F_D=0.05$ in Fig.~\ref{fig:3}(c,d).
The passive disks do not reach a pinned state with $\langle V\rangle \approx 0$
until $\phi_{\rm obs}>\phi_c=0.098$.
Figure~\ref{fig:3}(c) shows that $C_{L}$ is small for
low $\phi_{\rm obs}$ and increases to a
peak value of $C_L=0.5$ just below $\phi_{c}$. 
For $\phi_{\rm obs} > \phi_{c}$, $C_{L}$ decreases with increasing obstacle density.
In Fig.~\ref{fig:5}(a) we show a snapshot of the flowing state at
$\phi_{\rm obs} = 0.01178$.  Although no clusters appear, the disks tend 
to form one-dimensional (1D) flowing chains.
In the flowing state at $\phi_{\rm obs}=0.055$, illustrated in Fig.~\ref{fig:5}(b),
small clusters are beginning to appear.
The pinned cluster state
near the peak value of $C_{L}$ at $\phi_{\rm obs} = 0.1178$
is shown in Fig.~\ref{fig:2}(d).
Above the peak in $C_L$, the size of the clusters decreases with
increasing obstacle density and a
clogged state forms as shown in Fig.~\ref{fig:5}(c) for
$\phi_{\rm obs} = 0.14137$ and in Fig.~\ref{fig:5}(d) for
$\phi_{\rm obs} = 0.1963$, where the clusters have become quite small.

\begin{figure}
\includegraphics[width=3.5in]{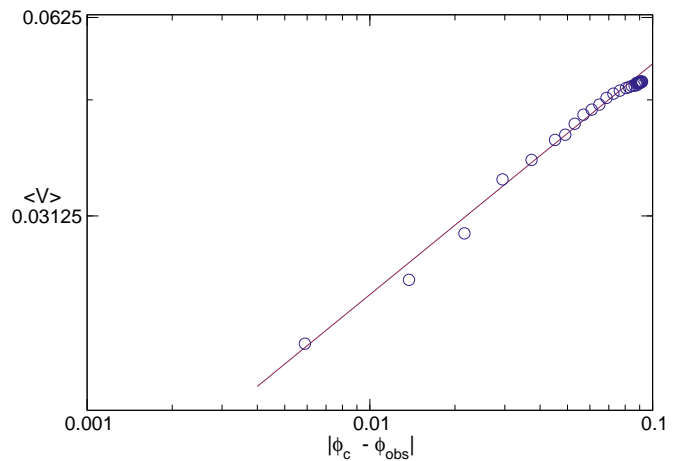}
\caption{The scaling of
  $V \propto |\phi_{c} - \phi_{\rm obs}|^\beta$
  for the passive disks in Fig.~\ref{fig:3}(d) with
  $\phi_c=0.098$ and 
  $\beta = 0.35$.
  }
\label{fig:6}
\end{figure}

The peak or divergence in $C_{L}$ for the passive disk system in Fig.~\ref{fig:3}(c)
suggests that the onset of complete clogging at
$\phi_{c}$ occurs at a critical point.
We have tried performing a power law fit
$C_{L} \propto |\phi_{c} - \phi_{\rm obs}|^\nu$
on either side of the divergence;
however, we find only a limited range for the fit resulting in strong
variations in the exponent.
We find more consistent scaling of
the average drift velocity as  $\phi_{c}$ is approached,
with $\langle V\rangle \propto |\phi_{c}  - \phi_{\rm obs}|^\beta$,
as shown in Fig.~\ref{fig:6} where $\beta = 0.35$.
We have studied
the critical clogging behavior of passive disks in
more depth in Ref.~\cite{14}, where we
find
a robust power law divergence in the
transient times near $\phi_{c}$
consistent with
an absorbing phase transition.
The focus of the present work is 
active ballistic jamming and we measure the passive disks for comparison.
Our results
indicate that the onset of pinned or clogged states
for the active ballistic disks is very different in nature from the
clogging of passive disks.
In particular,
we find only two phases for the passive disks and four phases for the
active ballistic disks.
The I$_{\rm fc}$-II transition marking the onset of a pinned state
for the active ballistic disks produces discontinuities in
both $\langle V\rangle$ and $C_{L}$,
consistent with a first order phase transition,
while for the passive disks, the onset of pinning or clogging
has the character of a second order phase
transition or a crossover phenomenon.

\begin{figure}
\includegraphics[width=3.5in]{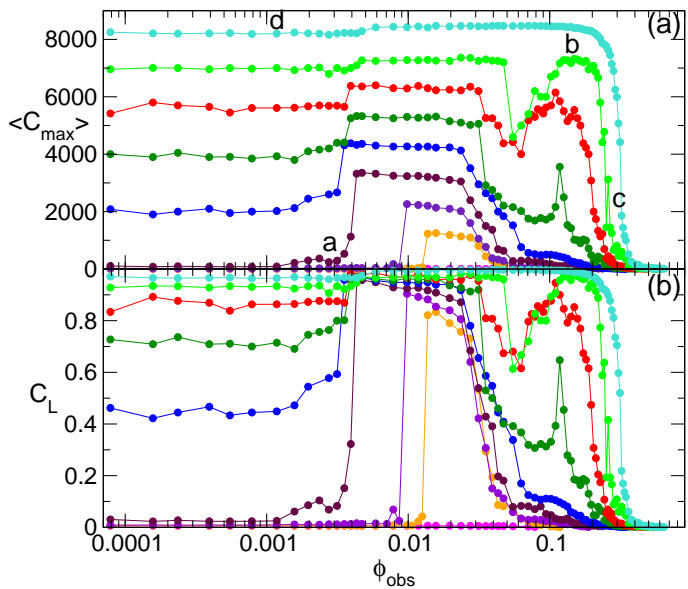}
\caption{ 
  (a) $\langle C_{\rm max}\rangle$, the average number of disks in the largest cluster,
  vs $\phi_{\rm obs}$ at $F_D=0.05$ for
  $\phi_{\rm tot}=0.668$ ($N_{\rm tot} = 8500$, turquoise),
  0.589 ($7500$, light green), 0.511 ($6500$, red), 0.432 ($5500$, dark green),
  0.354 ($4500$, blue), 0.275 ($3500$, maroon), 0.196 ($2500$, violet),
  0.118 ($1500$, orange), and 0.0786 ($1000$, magenta).
  (b) The corresponding $C_{L}$ vs $\phi_{\rm obs}$ curves.
  The I-II transition is associated with a jump or increase
  in $\langle C_{\rm max}\rangle$ and $C_{L}$,
  while at large $\phi_{\rm obs}$, the system enters a pinned disordered phase as
  indicated by the drop in $C_{\rm max}$ and $C_{L}$ to nearly zero.
  The labels {\bf a} to {\bf d} in (a) indicate the values of $\phi_{\rm obs}$ at which the
  images in Fig.~\ref{fig:8} were obtained.
}
\label{fig:7}
\end{figure}

\begin{figure}
\includegraphics[width=3.5in]{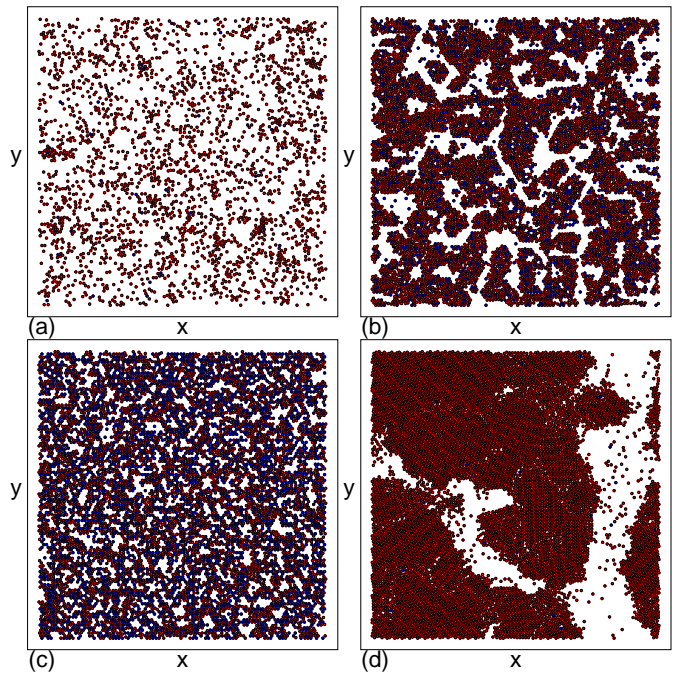}
\caption{
  The active ballistic disk positions (red circles) and obstacle locations (blue circles)
  for the system in Fig.~\ref{fig:7} obtained at the values
  of $\phi_{\rm tot}$ and $\phi_{\rm obs}$ marked by the letters
  {\bf a} to {\bf d} in Fig.~\ref{fig:7}.  (a) The drifting liquid
  for $\phi_{\rm tot} = 0.196$ and $\phi_{\rm obs} = 0.003926$.
  (b) The pinned gel phase containing a large-scale percolating cluster at
  $\phi_{\rm tot} = 0.589$ and $\phi_{\rm obs} = 0.1256$.
(c) The pinned disordered phase at $\phi_{\rm tot} = 0.589$ and $\phi_{\rm obs} = 0.283$.
  (d) The high density depinned fluctuating cluster state at
  $\phi_{\rm tot} = 0.668$ and $\phi_{\rm obs} = 0.000156$.
}
\label{fig:8}
\end{figure}

In Fig.~\ref{fig:7}(a) we plot $\langle C_{\rm max}\rangle$, the
average number of disks in the largest
cluster, versus $\phi_{\rm obs}$
for the active ballistic system at varied $\phi_{\rm tot}$ to highlight the evolution
of phases I through IV.
Figure~\ref{fig:7}(b) shows the collapse of the curves when the
same results are plotted in terms of $C_L$, the
average fraction of disks in the largest cluster, versus $\phi_{\rm obs}$.
For
$\phi_{\rm tot} \leq 0.275$,
phase I is
a uniform drifting liquid with $C_{L} <  0.04$, as illustrated
in Fig.~\ref{fig:8}(a) for $\phi_{\rm tot} = 0.196$
at $\phi_{\rm obs} = 0.003926$.
As $\phi_{\rm obs}$ increases,
there is a transition from the drifting liquid phase I to the pinned single cluster phase II,
as indicated by the large increase in $C_{L}$ to a value of $C_L=0.8$ or higher.
The obstacle density $\phi_{\rm obs}$ at which
the I-II transition occurs shifts upward as $\phi_{\rm tot}$ decreases,
and at the lowest values of $\phi_{\rm tot}$
that we consider,
the system always remains in phase I,
as shown for $\phi_{\rm tot} = 0.0786$ in Fig.~\ref{fig:7}.
We note that the maximum allowed value of
$\phi_{\rm obs}$
decreases with decreasing $\phi_{\rm tot}$
since it is bounded by the total disk density.
For $\phi_{\rm tot} \geq 0.35$, instead of the drifting liquid
phase I, we find the
depinned fluctuating cluster state I$_{\rm fc}$ as described previously,
and in all cases the
I-II and I$_{\rm fc}$-II transitions
are associated with a jump or increase in $C_{L}$.
The II-III transition
also shifts to
higher values of $\phi_{\rm obs}$ with increasing
$\phi_{\rm tot}$.

Within the pinned disordered phase IV in Fig.~\ref{fig:7},
an additional feature emerges for
$\phi_{\rm tot} = 0.432$
in the form of a peak
in $\langle C_{\rm max}\rangle $ and  $C_{L}$ near
$\phi_{\rm obs} = 0.115$.
This peak  grows in both height and extent with
increasing $\phi_{\rm tot}$.
At the onset of the pinned disordered phase IV, $C_L$ is low and the disks form a
small number of isolated clumps.  As $\phi_{\rm obs}$ increases, these
clumps break apart and the disks are spread more evenly over the substrate.
At higher overall disk densities $\phi_{\rm tot} \geq 0.432$, this produces
a percolation transition in which the broken clumps
merge to form a pinned gel or labyrinth state of the type illustrated
in Fig.~\ref{fig:8}(b)
at the peak in $\langle C_{\rm max}\rangle$ and $C_{L}$
for the $\phi_{\rm tot} = 0.589$ and $\phi_{\rm obs} = 0.1256$ system
in Fig.~\ref{fig:7}.
For higher $\phi_{\rm obs}$, the active
ballistic disks spread further apart
and the gel transforms to a pinned disordered state,
as shown
in Fig.~\ref{fig:8}(c) for the
$\phi_{\rm tot} = 0.589$ system from Fig.~\ref{fig:7} at $\phi_{\rm obs} = 0.283$.
The emergence of the intermediate
pinned gel
state is responsible for the
additional peak in $\langle C_{\rm max}\rangle$ and $C_L$.
At the highest value $\phi_{\rm tot}=0.668$ in Fig.~\ref{fig:7},
the II-III transition is lost and the $\langle C_{\rm max}\rangle$ and
$C_L$ curves become nearly featureless below the transition to the
pinned disordered phase IV.
When the total disk density is high,
motion in the depinned fluctuating cluster phase I$_{\rm fc}$
becomes less intermittent and the steady state value of $C_L$ increases
to $C_{L} > 0.95$.
The high density depinned fluctuating cluster state is
illustrated in Fig.~\ref{fig:8}(d) for $\phi_{\rm tot} = 0.668$ and $\phi_{\rm obs} = 0.00156$.

\begin{figure}
\includegraphics[width=3.5in]{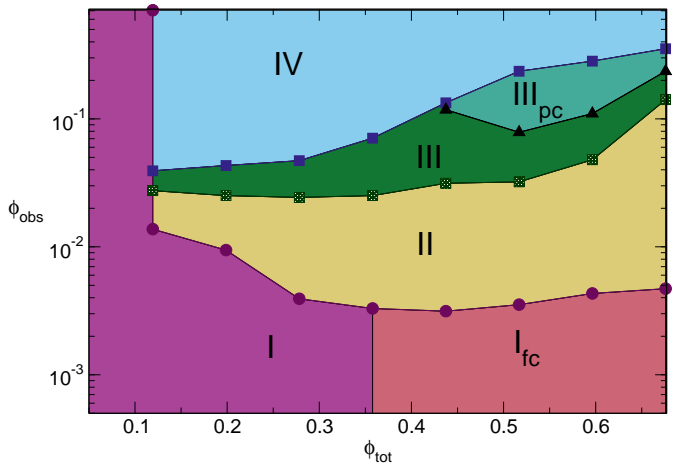}
\caption{
  The locations of the phases in Fig.~\ref{fig:7} as a function of
  $\phi_{\rm obs}$ vs $\phi_{\rm tot}$.
  Phase I (magenta) is the drifting liquid state;
  phase I$_{\rm fc}$ (pink) is the depinned fluctuating clump state;
  phase II (yellow) is the pinned single cluster state;
  phase III (dark green) is the pinned multicluster state;
  phase III$_{\rm pc}$ (light green) is the
  pinned gel state;
  and phase IV (blue) is the pinned disordered state.
}
\label{fig:9}
\end{figure}

Using the results in Fig.~\ref{fig:7}, we can construct a phase diagram
showing the evolution of the different phases as a function
of $\phi_{\rm obs}$ versus $\phi_{\rm tot}$,
as shown in Fig.~\ref{fig:9}.
The unpinned state is a drifting liquid (phase I) for low $\phi_{\rm tot}$
and a depinned fluctuating clump state (phase I$_{\rm fc}$) for
high $\phi_{\rm tot}$.
Phase II is the pinned single clump state and
phase III is the pinned multiclump state.
For large $\phi_{\rm tot}$ we find a window of phase III$_{\rm pc}$,
the pinned gel state, at $\phi_{\rm obs}$ values
above phase III.  
Phase IV is the pinned disordered state.
The features in the $C_{L}$ and $\langle V\rangle$
curves indicate that the I-II and I$_{\rm fc}$-II transitions
are first order in nature,
while the II-III and III-IV transitions are continuous or show crossover behavior.
In Ref.~\cite{14} we perform a detailed study of
the behavior of passive disks as a function of $\phi_{\rm obs}$ versus $\phi_{\rm tot}$,
where we find that the pinned phase appears at much higher values of $\phi_{\rm obs}$
than in the active ballistic system.
We note that there may be additional phases in the active
ballistic system at values of $\phi_{\rm tot}$
higher than those shown in Fig.~\ref{fig:9}, particularly upon
approaching $\phi_{\rm tot}=0.9$ where the system crystallizes into a
close-packed lattice.

\begin{figure}
\includegraphics[width=3.5in]{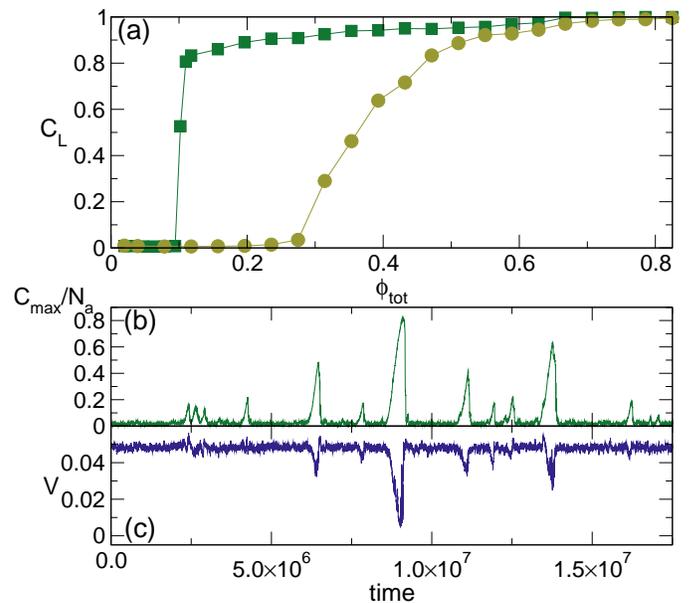}
\caption{
  (a) $C_{L}$
  vs $\phi_{\rm tot}$ for active ballistic disks
  with $F_D=0.05$.
  Light green circles: In the obstacle-free system $\phi_{\rm obs}=0$,
  clustering begins near $\phi_{\rm tot} = 0.28$.
  Dark green squares: A system with
  $\phi_{\rm obs} = 0.00157$, showing that
  inclusion of a small number of obstacles can stabilize a cluster state
  at much lower densities $\phi_{\rm tot} \approx 0.1$.
  (c) $C_{\rm max}/N_a$ and (d) instantaneous velocity $V$
  versus time for a system with $\phi_{\rm tot}=0.196$ and $\phi_{\rm obs}=0.00785$
  that remains in phase I
  but is near the I-II transition.
  Drops in $V$ indicate the
  temporary formation of a pinned cluster as shown by the
  corresponding jumps in $C_{\rm max}/N_a$.
}
\label{fig:10}
\end{figure}
 
In nonballistic active particle systems that undergo driven diffusion or have
run-and-tumble motion with finite $\tau_r$,
a phase separated state appears in the absence of quenched disorder as
a function of
disk density $\phi_{\rm tot}$ and activity.
Typically there is a density $\phi_{\rm tot}^{\rm min}$ below which phase
separation does not occur.
For the ballistic active matter system we consider, in the absence of obstacles a
phase separated state appears only for
$\phi_{\rm tot} \geq 0.35$.
The introduction of obstacles makes it possible for a cluster state to
nucleate at much smaller values of $\phi_{\rm tot}$.
For $F_{D} = 0.05$, we find clustering at densities as small as
$\phi_{\rm tot} \approx  0.1$.
To illustrate this more clearly,
in Fig.~\ref{fig:10} we plot
$C_{L}$ versus $\phi_{\rm tot}$ for
systems with $\phi_{\rm obs} = 0$ and
$\phi_{\rm obs} = 0.00157$.
In the obstacle-free system, the cluster size begins to increase
near $\phi_{\rm tot} = 0.28$, reaching a value of
$C_{L} = 0.8$ at $\phi_{\rm tot} = 0.47$.
In contrast, adding a small number of obstacles shifts
the
onset of clustering
down to
$\phi_{\rm tot} = 0.095$,
and $C_{L}$ reaches a value of $C_{L} = 0.8$ at $\phi_{\rm tot} = 0.12$.
This indicates that the obstacles are responsible for nucleating stable clusters
over the range
$0.12 < \phi_{\rm tot} < 0.28$.
The disorder-induced cluster state can be viewed
in terms of a wetting phenomenon
where the active particles accumulate not along walls \cite{71}  but next to the obstacles.

The ability of a cluster to form at low obstacle densities
also depends on $F_{D}$.
As $F_D$ decreases,
the transition from the drifting liquid phase I
to the pinned single cluster phase II occurs at
lower
$\phi_{\rm obs}$.
In general, even when $F_D$ is too large to stabilize phase II for a given value
of $\phi_{\rm obs}$,
a transient pinned cluster can still form on a temporary basis.
An example of this
appears in Fig.~\ref{fig:10}(b,c) where we plot
$C_{\rm max}/N_a$ and instantaneous velocity $V$ versus
time
for a system with $\phi_{\rm tot} = 0.196$ and $\phi_{\rm obs}  = 0.00785$,
which is close to the I-II transition on the phase I side.
There are a series of dips in $V$
correlated with jumps in $C_{\rm max}/N_a$ which arise
when a pinned cluster forms, reducing the velocity temporarily.
The cluster quickly breaks apart, restoring $V$ and $C_{\rm max}/N_a$ to their
steady state average values.

\section{Depinning and Drive Dependence}

\begin{figure}
\includegraphics[width=3.5in]{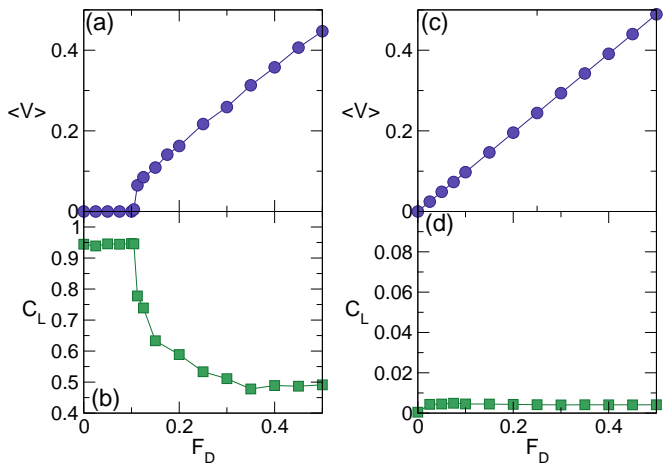}
\caption{
  (a) $\langle V\rangle$ and (b) $C_{L}$ vs $F_{D}$
  for active ballistic disks with
  $\phi_{\rm tot} = 0.3534$ and $\phi_{\rm obs} = 0.00178$.
  At $F_D=0.05$ the system is in phase II, the pinned single cluster state,
  and at $F_D=F_c=0.1$ it depins and enters phase I$_{\rm fc}$.
  (c) $\langle V\rangle$ and (d) $C_L$ vs $F_D$
  for passive ballistic disks at the same values of $\phi_{\rm tot}$ and
  $\phi_{\rm obs}$.
  There is no depinning threshold and clustering does not occur.
}
\label{fig:11}
\end{figure}

We next consider the effect of changing $F_{D}$ in order to
construct velocity-force ($v-f$) curves and measure
the depinning threshold $F_{c}$.
We compare the depinning of the active ballistic disks to that of
passive disks.
In Fig.~\ref{fig:11}(a,b)
we plot $\langle V\rangle$ and $C_{L}$ versus $F_{D}$
for an active ballistic system with $\phi_{\rm tot} = 0.3534$
and $\phi_{\rm obs} = 0.00178$, which is in the pinned single cluster phase II
at $F_{D} = 0.05$.
We find a drive-induced depinning transition at $F_D=F_c=0.1$
from phase II to phase I$_{\rm fc}$, as shown by
the sharp drop in  $C_{L}$ that coincides with
the onset of a linear increase in $\langle V\rangle$ with increasing $F_D$.
In Fig.~\ref{fig:11}(c,d) we show $\langle V\rangle$ and $C_L$ versus $F_D$ for
a passive disk system with the same values of $\phi_{\rm tot}$ and $\phi_{\rm obs}$.
There is no depinning threshold and $C_{L} < 0.005$ for all values of $F_{D}$,
indicating a complete lack of clustering.

\begin{figure}
\includegraphics[width=3.5in]{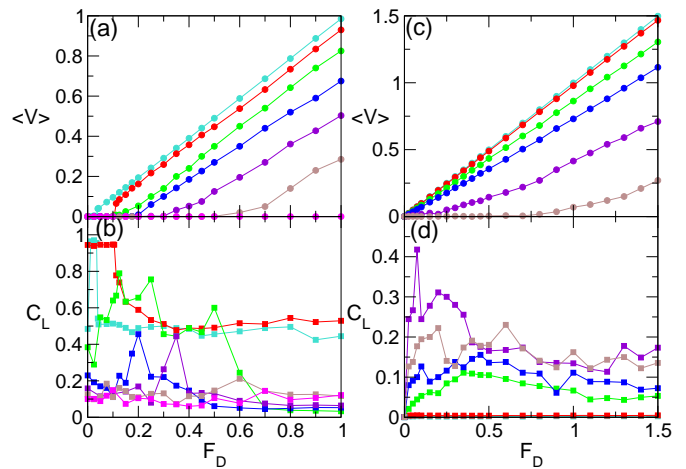}
\caption{ 
  (a) $\langle V\rangle$ and (b) $C_{L}$ vs $F_{D}$
  for active ballistic disks with $\phi_{\rm tot} = 0.3534$ at
  $\phi_{\rm obs} = 0.00235$ (light blue),
  $0.01178$ (red), $0.03927$ (green), $0.0628$ (dark blue),
$0.0785$ (violet), $0.0942$ (brown), and $0.1256$ (magenta).
  (c) $\langle V\rangle$ and (d) $C_{L}$ vs $F_{D}$ for
  passive disks with $\phi_{\rm tot} = 0.3534$ at
  $\phi_{\rm obs} = 0.00235$ (light blue),
  $0.01178$ (red), $0.03927$ (green), $0.0628$ (dark blue),
  $0.094274$ (violet), and $0.1267$ (brown),
  showing that a finite depinning threshold does not appear until
$\phi_{\rm obs} > 0.094274$.
}
\label{fig:12}
\end{figure}

In Fig.~\ref{fig:12}(a,b) we plot
$\langle V\rangle$ and $C_{L}$ versus $F_{D}$ for the active ballistic disks with
$\phi_{\rm tot} = 0.3534$ at
$\phi_{\rm obs} = 0.00235$ 
to $0.1256$.
We find that the II-I$_{\rm fc}$ transition,
corresponding to the depinning threshold,
drops to lower values of $F_D$ as $\phi_{\rm obs}$ decreases,
as indicated most clearly by the $\phi_{\rm obs}=0.00235$ curve in Fig.~\ref{fig:12}(a)
which has $F_c=0.035$.
The II-I$_{\rm fc}$ depinning transition
is generally quite sharp, and the system goes directly from
the pinned state to a
fully flowing state without
passing through a regime in which moving and
pinned active particles coexist.
In contrast, the depinning transition separating phases III and IV for
$\phi_{\rm obs} \geq 0.03927$
is smooth or continuous and is plastic in nature, so that above depinning
only a portion of the active disks are flowing while the other portion remains pinned.
In studies of depinning in other systems such as
superconducting vortices or colloidal particles
moving over quenched disorder,
elastic depinning is associated with a sharp transition in 
the $v-f$ curve
and a scaling of $V \propto (F_{D} -F_{c})^\alpha$ with $\alpha < 1.0$.
Plastic depinning is accompanied by 
an extensive nonlinear regime in the $v-f$ curves with
$\alpha >1.0$.
In our active ballistic disk system, the resolution of the
$v-f$ curves is not high enough to perform a
scaling analysis;
however, the qualitative change in the $v-f$ curve
at the $I_{\rm fc}$-II transition is consistent with
an elastic or collective depinning at low $\phi_{\rm obs}$
crossing over to plastic depinning for higher $\phi_{\rm obs}$.
For the III-IV depinning
transition, there is generally a small peak in
$C_{L}$ at $F_{c}$
produced when the disks
start to accumulate behind the obstacles,
and in all cases there is an overall drop in
$C_{L}$ at higher values of $F_{D}$ in the moving phase. 
For $\phi_{\rm obs} = 0.1256$, depinning does not
occur until $F_D=F_{c} = 1.35$.

In Fig.~\ref{fig:12}(c,d) we plot $\langle V\rangle$ and $C_{L}$
versus $F_{D}$ for the passive disks with
$\phi_{\rm tot} = 0.3534$ at
$\phi_{\rm obs} = 0.00235$
to $0.1267$.
There is no finite depinning threshold for
$\phi_{\rm obs} \leq 0.094274$.
We find an extended
regime of nonlinear flow for
$\phi_{\rm obs} > 0.0628$
associated with a coexistence of flowing and clogged disks. 
In Fig.~\ref{fig:12}(d),
$C_{L} < 0.01$ at all $F_D$ for $\phi_{\rm obs} = 0.00235$ and
$\phi_{\rm obs}=0.01178$, while for higher $\phi_{\rm obs}$ there is
generally a decrease in $C_{L}$ with increasing $F_D$ for $F_D>0.3$.
The maximum value of $C_{L}$ occurs for $\phi_{\rm obs} = 0.094274$,
which is just below the obstacle density at which
a finite depinning threshold first appears.  These results show  that
for varied $F_{D}$, the active ballistic disks have a
much higher susceptibility to becoming pinned
than the passive disks.

\begin{figure}
\includegraphics[width=3.5in]{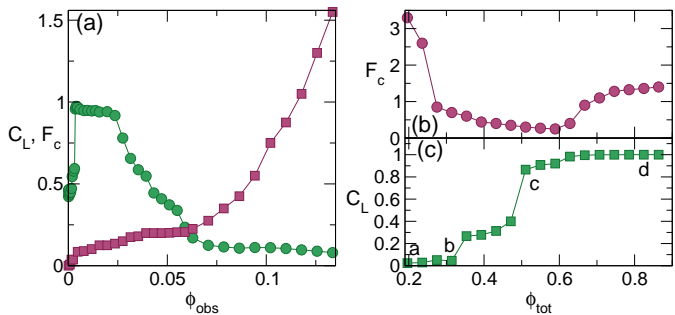}
\caption{
  (a) $F_{c}$ (red squares) and the value of $C_{L}$ at $F_D=0.05$ (green circles) vs
  $\phi_{\rm obs}$ for the active ballistic disks at $\phi_{\rm tot} = 0.3534$.
  The depinning threshold remains finite
  down to $\phi_{\rm obs} = 0.0008$,
  and there is an increase in $F_{c}$ at the onset of phase IV.
  (b) $F_{c}$ and (c) the value of $C_{L}$ at $F_D=0.05$  vs
  $\phi_{\rm tot}$ for $\phi_{\rm obs} = 0.09427$.
  The behavior of $F_c$ is nonmonotonic, and
  $F_c$ increases with increasing $\phi_{\rm tot}$ for $\phi_{\rm tot}>0.6$ when
  $C_L=1$, indicating the formation of a jammed state.
  The labels {\bf a} to {\bf d} in (c) indicate the values of $\phi_{\rm tot}$ at which the
  images in Fig.~\ref{fig:14} were obtained.
}
\label{fig:13}
\end{figure}

In Fig.~\ref{fig:13}(a) we plot
the evolution of the
depinning threshold $F_{c}$
along with the value of $C_{L}$ at $F_D=0.05$
versus $\phi_{\rm obs}$ for the active ballistic disks
at $\phi_{\rm tot} = 0.3534$.
There is a sharp increase from $F_c=0$ at $\phi_{\rm obs}=0$ to
$F_c=0.007$ at $\phi_{\rm obs}=0.0008$, the lowest nonzero obstacle density
we considered, for which
the ratio of obstacles to
active particles is $N_{a}/N_{\rm obs} =  440$.
For $ 0.0045 < \phi_{\rm obs} < 0.039$, there is
a more gradual  linear increase of $F_{c}$ with increasing $\phi_{\rm obs}$
over the range of phase II depinning through half of the phase III depinning.
This is followed 
by a regime of roughly constant $F_{c}$
for $0.039 < \phi_{\rm obs} < 0.065$
in the second half of phase III depinning, while
in phase IV for $\phi_{\rm obs} > 0.65$,
there is a rapid increase in $F_{c}$ which coincides with a drop in $C_{L}$.
Plastic depinning appears in phase IV, and the rapid increase in $F_c$ in this
regime is reminiscent of
the ``peak effect'' phenomenon observed for the depinning of  superconducting vortices.
At low disorder strength, the vortices form a crystal that that can be
collectively pinned by a small number of pinning sites; however, when the
pinning strength is increased, the crystalline structure breaks apart,
the vortex structure becomes amorphous, and there is a pronounced increase
in the depinning force that is much larger than
what would be expected from the increase in
the pinning strength
\cite{2,3}. In the active ballistic disk system, the clusters have
considerable crystalline order, and when the amount of disorder is increased
by raising the number of obstacles,
the large clusters break up into smaller
clusters that can be pinned more easily, leading to the increase in $F_c$.

\begin{figure}
\includegraphics[width=3.5in]{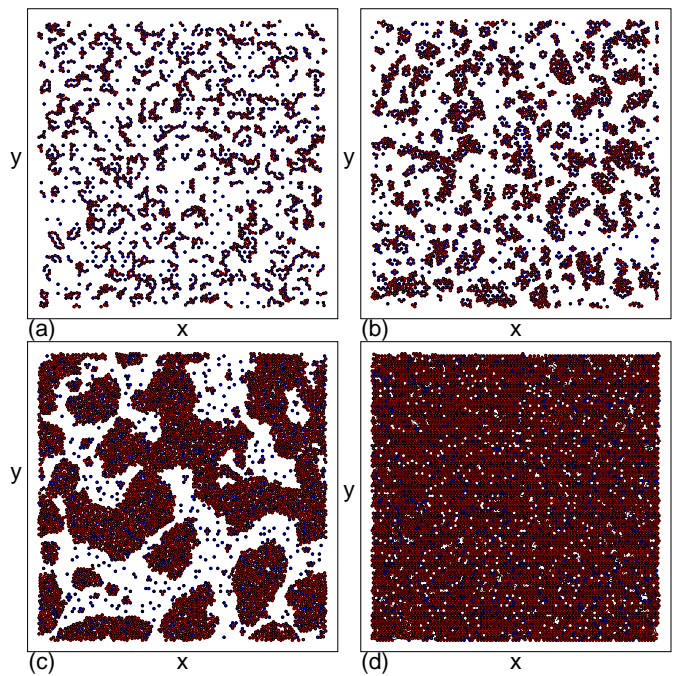}
\caption{
  The active ballistic disk positions (red circles) and obstacle locations
  (blue circles) for the system in Fig.~\ref{fig:13}(b,c) with
  $\phi_{\rm obs} = 0.09427$
  obtained at the values
  of $\phi_{\rm tot}$ marked by the letters {\bf a} to {\bf d} in Fig.~\ref{fig:13}(c).
  (a) The pinned liquid at $\phi_{\rm tot} = 0.1963$.
  (b) A pinned weakly clustered state at
  $\phi_{\rm obs} = 0.275$ (phase IV).
  (c) A pinned gel state at $\phi_{\rm tot} = 0.51$ (phase III$_{\rm pc}$).
  (d) A jammed solid state at $\phi_{\rm tot} = 0.8246$.
}
\label{fig:14}
\end{figure}

In Fig.~\ref{fig:13}(b,c) we plot $F_{c}$ and
the value of $C_{L}$ at $F_D=0.05$ versus
$\phi_{\rm tot}$ for the active ballistic disks
at $\phi_{\rm obs} = 0.09427$.
For low $\phi_{\rm tot} < 0.35$, a disordered
but strongly pinned phase appears, as indicated by the large $F_c$ and the
low
$C_{L} < 0.1$.
In Fig.~\ref{fig:14}(a) we illustrate the disk configuration
at $\phi_{\rm tot} = 0.1963$ where a pinned liquid with small local disk clusters forms.
As $\phi_{\rm tot}$ increases,
$F_{c}$ decreases and a pinned weakly clustered state emerges,
as shown
in Fig.~\ref{fig:14}(b) at $\phi_{\rm tot} = 0.275$.
Since the number of obstacles is fixed,
as $\phi_{\rm tot}$ increases, each obstacle must restrain a larger number of
mobile disks, causing $F_c$ to decrease with increasing $\phi_{\rm tot}$.
In Fig.~\ref{fig:14}(c)
we plot the disk configuration at $\phi_{\rm tot} = 0.51$, where the system
forms a pinned gel phase with low $F_{c}$.
Figure~\ref{fig:13}(c,d) shows
a local minimum in $F_{c}$
near $\phi_{\rm tot} = 0.628$,
where $C_{L} = 0.92$.
For $\phi_{\rm tot} > 0.628$,
$F_{c}$ begins increasing with increasing $\phi_{\rm tot}$
and $C_{L}$ approaches $C_L=1$
as the disks assemble into
a single jammed solid  packing,
as shown  in Fig.~\ref{fig:14}(d) at $\phi_{\rm tot} = 0.8246$.

\begin{figure}
\includegraphics[width=3.5in]{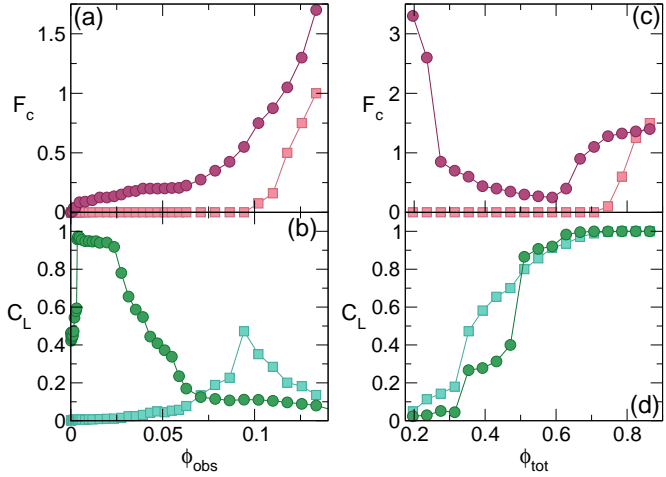}
\caption{ (a) $F_{c}$ and (b)
  the value of $C_{L}$ at $F_D=0.05$
  vs $\phi_{\rm obs}$ at $\phi_{\rm tot}=0.3534$
  for the active ballistic disks (circles) and passive disks
  (squares).
  (c) $F_{c}$ and (d) the value of $C_{L}$ at $F_D=0.05$
  vs $\phi_{\rm tot}$ at $\phi_{\rm obs}=0.094$ 
  for the active ballistic disks (circles) and passive disks
  (squares).
  At high $\phi_{\rm tot}$,
  both the active and passive disks undergo a transition into a jammed state.
}
\label{fig:15}
\end{figure}

\begin{figure}
\includegraphics[width=3.5in]{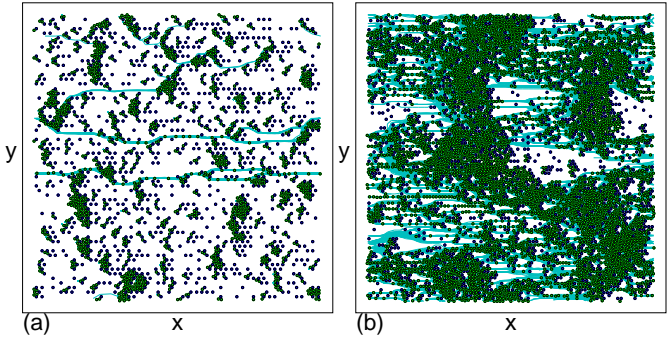}
\caption{ The passive disk positions (green circles),
  obstacle locations (blue circles), and disk trajectories
  for the system in Fig.~\ref{fig:15}(c,d) 
  (a) At $\phi_{\rm tot}= 0.196$, 1D flow channels coexist with pinned disks.
  (b) At $\phi_{\rm tot} = 0.4712$, collective interactions between disks
  cause a larger fraction of the disks to flow.
  }
\label{fig:16}
\end{figure}

In Fig.~\ref{fig:15}(a,b) we plot
$F_{c}$ and the value of $C_{L}$ at $F_D=0.05$
versus $\phi_{\rm obs}$ at $\phi_{\rm tot}=0.3534$
for active ballistic disks and passive disks.
The depinning threshold remains finite in the active system
for $\phi_{\rm obs} \geq 0.0008$, whereas in the passive system
the depinning threshold drops to zero when $\phi_{\rm obs} \leq 0.094$.
This indicates that 100 times fewer obstacles are needed to pin the
active system compared to the passive system.
In addition, the depinning threshold for the active system 
is always higher than that of the passive system. 
In Fig.~\ref{fig:15}(c,d) we show
$F_{c}$ and the value of $C_{L}$ at $F_D=0.05$ versus $\phi_{\rm tot}$ at
$\phi_{\rm obs} = 0.094$ for the active ballistic and passive disks.
Here the depinning threshold for the passive disks does not become
finite
until $\phi_{\rm tot} > 0.7$, the density above which $C_L$ increases to $C_L=1.0$.
The high density active ballistic and passive disk states are similar in nature and are
dominated
by the formation of a jammed solid state.
Even when the depinning threshold in the passive disk system is zero,
the $v-f$ curves can show strongly nonlinear behavior,
and a large fraction of the passive disks remain pinned or immobile
for $\phi_{\rm tot} < 0.35$, the same total disk density at which
the active ballistic disks exhibit a large increase in $F_{c}$. 
In Fig.~\ref{fig:16}(a) we show the
passive disk locations and trajectories
at $\phi_{\rm tot} = 0.196$
and $F_{D} = 0.05$.
A portion of the disks
move in 1D filamentary channels while the remaining disks are pinned.
These filamentary channels
persist down to $F_{D} = 0.0$ in the passive disks, but
are generally absent for active ballistic disks.
In the passive disk system,
as $\phi_{\rm tot}$ increases, cooperative interactions between the mobile
disks reduce the overall trapping and lead to a higher fraction of
flowing disks, as illustrated
in Fig.~\ref{fig:16}(b) at $F_{D} = 0.05$ and $\phi_{\rm tot} = 0.4712$.
As $\phi_{\rm tot}$ further increases,
the system approaches a jammed solid with a finite depinning threshold.

\section{Finite Run Time Active Disks}

\begin{figure}
\includegraphics[width=3.5in]{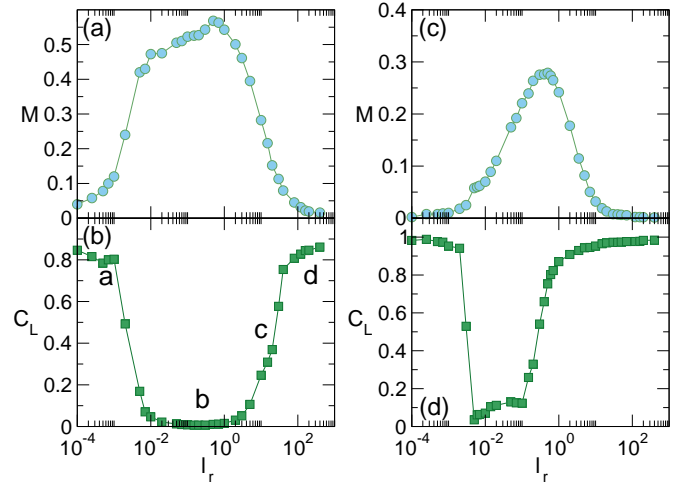}
\caption{ (a) The disk mobility $M$
  and (b) $C_{L}$ vs 
  run length $l_{r}$ for finite run time active disks at
  $F_D=0.05$, $\phi_{\rm tot} = 0.51$ and $\phi_{\rm obs} = 0.14137$.
  The low $l_{r}$ behavior is similar to
  that found in the passive disk limit while
  the high $l_{r}$ behavior is similar to that found in the active ballistic limit.
  Between these two limits the disks form a uniform
  density, highly mobile liquid state.
  The labels {\bf a} to {\bf d} in (b) indicate the values
  of $l_{r}$ at which the images in
  Fig.~\ref{fig:18} were obtained.
  (c) $M$ and (d) $C_L$ vs $l_r$
  for finite run time active disks at $\phi_{\rm tot} = 0.667$
  and $\phi_{\rm obs}=0.14137$.
}
\label{fig:17}
\end{figure}

\begin{figure}
\includegraphics[width=3.5in]{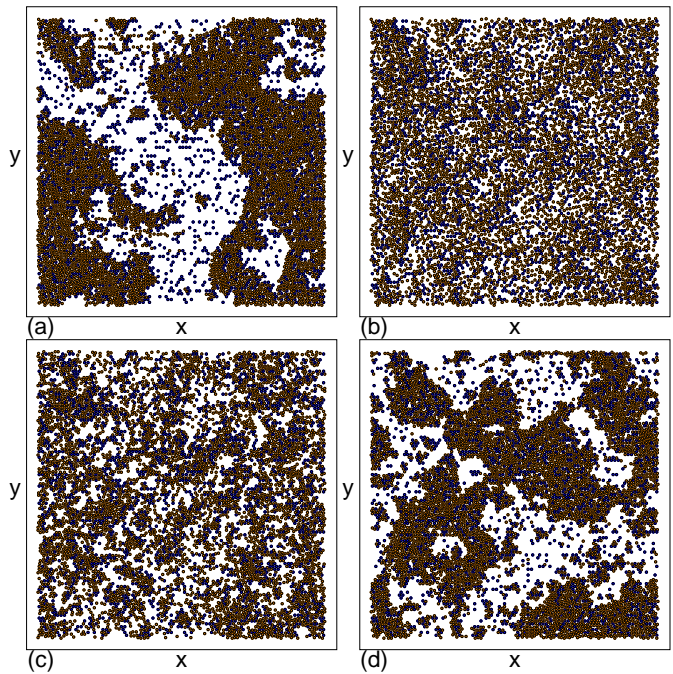}
\caption{ The finite run time active disk locations (orange circles) and obstacle
  locations (blue circles) for a system with
  $\phi_{\rm tot} = 0.51$ and $\phi_{\rm obs} = 0.14137$
  obtained at the values of $l_r$ marked by the letters {\bf a} to {\bf d} in
  Fig.~\ref{fig:17}(b).
  (a) The low mobility clogged state at $l_{r} = 0.001$.
  (b) The high mobility uniform liquid at $l_{r} = 2.0$.
  (c) The actively clogged state at $l_{r} =10$.
(d) The actively clogged state at $l_{r}= 200$.
}
\label{fig:18}
\end{figure}

We next show how the passive disk pinning and
active ballistic disk pinning limits can be connected to each other by 
considering run-and-tumble particles where we gradually increase
the running time from close to zero, which is the passive limit, to large
values which approach the ballistic limit.
In Fig.~\ref{fig:17}(a,b) we plot
the mobility 
$M = \langle V\rangle/\langle V_{0}\rangle$ per disk
and $C_{L}$ versus run length $l_{r}$
for a finite run time active disk system at
$\phi_{\rm tot} = 0.51$,  $\phi_{\rm obs} = 0.14137$, and $F_{D} = 0.05$.
Here $\langle V_0\rangle$ is the average drift velocity of an individual
disk in the absence of obstacles or other disks, so for $F_D=0.05$,
$\langle V_0\rangle=0.05$, and in the obstacle-free limit,
$M = 1.0$.
The disk dynamics are the same as those described in Eq.~1 except the
running time $\tau_r$ is now finite.
For convenience, we characterize the activity level in the system
at a fixed $|{\bf F}_{m}|$ in terms of the run length 
$l_{r} = |{\bf F}_{m}|\tau_{r}$,
so that large $l_{r}$ corresponds
to large $\tau_{r}$.
In the passive limit, $l_r=0$, while in the ballistic limit, $l_r$ is infinite.
For the parameters we consider in this section,
the system reaches a completely pinned state in both the passive and ballistic limits.
In Fig.~\ref{fig:17}(a) at low $l_{r}$,
$\langle M\rangle$ is small, indicating that a clogged state has formed that is
similar to the passive disk clogged state which appears at $l_r=0$.
At the same time,
$C_L>0.8$
in Fig.~\ref{fig:17}(b),
indicating the
formation of a large cluster.
In Fig.~\ref{fig:18}(a) we plot the
disk configurations for $l_{r} = 0.001$, where the clogged state has highly
heterogeneous local disk density.
For $ 0.01 < l_{r} < 5$, an easily flowing liquid appears,
as indicated by the increase in $M$ and the drop in $C_L$ to 
$C_{L} < 0.01$.  We illustrate the flowing liquid at $l_r=2.0$ in
Fig.~\ref{fig:18}(b), where the disk density is uniform and clustering
behavior is lost.
As $l_{r}$ increases, $M$ decreases once self-clustering of the disks
begins to occur,
as shown in 
Fig.~\ref{fig:18}(c) for $l_r=10$.
At large $l_{r}$, $C_{L}$ approaches $C_L=0.9$
and $M$ drops to a low value.  In this regime, 
the image of the $l_r=200$ system in Fig.~\ref{fig:18}(d) indicates that
clustering similar to that found in the active ballistic pinned
state is present.
We note that $M>0$ for any finite $l_{r}$
since there is always a 
chance that the activity can unpin a fraction of the disks even
when $l_r$ becomes large.
The motion in the finite but large $l_{r}$ regime
becomes highly intermittent and exhibits avalanche-like
fluctuations, as has been described in
detail elsewhere \cite{62}.
In Fig.~\ref{fig:17}(c,d) we plot $M$ and $C_L$ versus $l_r$ for
the run-and-tumble disks at
a higher $\phi_{\rm tot} = 0.667$.
We find the same trend in which clustering occurs at
both small and large $l_{r}$,
with a correspondingly low value of $M$,
while for intermediate $l_r$, the clustering disappears,
$C_L$ is low, and $M$ is high.
The maximum value of $M$ is lower at $\phi_{\rm tot}=0.667$
than at $\phi_{\rm tot}=0.51$
due to the crowding effect that appears
at higher disk densities.

\begin{figure}
\includegraphics[width=3.5in]{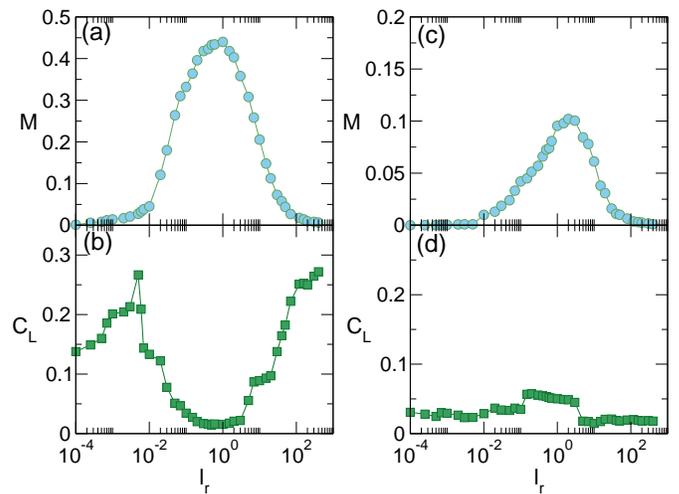}
\caption{ (a) The disk mobility $M$ and
  (b) $C_{L}$ vs run length $l_{r}$
  at  $F_{D} = 0.05$,
$\phi_{\rm tot} = 0.3543$ and $\phi_{\rm obs} = 0.14137$ for finite run time active disks.
  (c) $M$ and (d) $C_L$ vs $l_r$ at $\phi_{\rm tot} = 0.2356$ and $\phi_{\rm obs} = 0.14137$
  for finite run time active disks, where the system is always in the disordered regime.
}
\label{fig:19}
\end{figure}

In Fig.~\ref{fig:19}(a,b) we plot $M$ and $C_{L}$
for finite run time active disks
at $\phi_{\rm tot}=0.3534$ and $\phi_{\rm obs} = 0.1413$.
We find the same trend as in Fig.~\ref{fig:17} where
high values of $C_{L}$ are associated with low values of $M$;
however, at this lower total disk density
$\phi_{\rm tot}$, the maximum value of $C_{L}$ is reduced.
A peak in $C_L$ near $l_r=0.005$ indicates the appearance of additional
clustering just before the activity becomes large enough to liquefy the system
and increase $M$.
This suggests that the
activity becomes strong enough to cause
untrapping of individual disks at lower $l_{r}$ but that the untrapped disks
then pool into clusters that are too large to be broken apart by individual disks
until $l_r$ increases, at which point
the clumps break apart and the system flows
in a liquid state.
In Fig.~\ref{fig:19}(c,d) we plot $M$ and $C_L$ versus $l_r$
for finite run length disks at $\phi_{\rm tot} = 0.2356$ and $\phi_{\rm obs} = 0.1413$.
At this low total disk density, little clustering occurs and $C_{L}$ is always small.
We still observe a peak in $M$ at intermediate $l_r$ values;
however, $M$ is relatively low overall, reaching a maximum value of only $M=0.1$.
This result emphasizes 
the role of collective disk-disk interactions in liquefying the system and increasing
the mobility for intermediate $l_r$.

Although we consider only monodisperse disks in this work, the introduction of
obstacles causes the clogged states to exhibit
a considerable amount of structural disorder,
similar to that found in
jammed states
for bidisperse or 
completely amorphous systems \cite{65,67,68}.
For passive disks
at low obstacle densities, previous studies have 
shown that
jamming
occurs near the
obstacle-free jamming density of $\phi_{j}$ or point $J$, but that the
jamming transition shifts to lower disk densities as the obstacle density
increases
\cite{67,69}.
When the obstacle density is 
high enough, the behavior of the system changes and instead of
a uniform jammed state, the passive disks assemble into a
clogged state with strongly heterogeneous local disk density \cite{14}.
For active disks we find three limiting regimes of behavior.
At high disk densities, the effect of the activity becomes negligible and the behavior
is similar to that found in the passive high disk density limit,
which is controlled by jamming near point J in amorphous systems or
crystallization at a density $\phi_{\rm tot}=0.9$ for monodisperse disks.
The second limiting regime appears for low activity and intermediate disk densities,
where the active disks form a clogged configuration similar to that found for
the clogging of passive disks.
The third limiting regime, consisting of the actively pinned or clogged states that
appear for active ballistic disks or for disks with finite but large $l_r$,
is unique to the active disks and does not appear in
passive disk systems.  
This regime extends over a wide range
of obstacle densities and can assume the form of phase II, III, or IV, as described
in this work.
Our results suggest that in addition to the density, temperature, and load axes
on the jamming phase diagram  \cite{65,66},
there could be two additional axes, the activity level and the obstacle density or
disorder level,
that produce jammed states.

In future work for both the passive and active
disk systems, it would be interesting to investigate how different the
pinned phases are.
For example,
at high disk densities near the jamming transition,
it is likely that the system is highly stable
to perturbations since there are fewer available
degrees of freedom.
Similarly, in a clogged state with high obstacle densities, a perturbed system
is likely to fall back into the same or a similar clogged state.
On the other hand, in phases II or III,
a small perturbation could readily break up one or more of the clusters,
permitting the system to flow again,
so that although the active systems are more
susceptible to clogging,
the clogged state they reach may be more fragile than
that formed by passive disks.     

\section{Summary}
We have examined the pinning and clogging behaviors of active
run-and-tumble disks
in the ballistic limit driven through 
an array of obstacles.
As a function of increasing obstacle density,
we find four generic phases:
an unpinned fluctuating cluster phase,
a pinned single cluster phase in which a small number of obstacles can
pin a large number of active disks,
a pinned multiclump
phase, and
a pinned disordered phase.
We find that in contrast to passive disks,
the active ballistic disks can reach a pinned state at
relatively low obstacle densities.
Within the pinned disordered phase, as the density of active ballistic disks increases,
a pinned gel state or labyrinth pattern appears. 
By constructing velocity-force curves we
find that the active ballistic disks exhibit a
finite depinning threshold.
As the obstacle density increases, there is a transition from
collective depinning of the pinned clusters to
plastic depinning of the disordered pinned states
that is associated with
a large increase in the depinning threshold.
As a function of total disk density,
the depinning threshold for the active ballistic disks is nonmonotonic,
dropping at intermediate disk densities when collective disk-disk interactions
reduce the threshold, but rising at high disk densities as the system approaches
a jammed or crystalline state.
In contrast, for passive disks the depinning threshold is only finite
at high disk densities, while at lower disk densities filamentary channels of
disk flow form that are absent in the active disk system.
For both the active and passive disks, the pinned or clogged states are 
phase separated; however, the phase separation appears
at a much lower obstacle density for the active disks.
Finite run time active
disks provide a connection between the
active ballistic and passive disk systems, and
exhibit a low mobility phase separated state for short run times
in the passive limit as well as for long run times
in the active ballistic limit.
At intermediate run times, 
the active disks form an easily flowing uniform liquid
with reduced clustering, and there is
an optimal level of activity that maximizes the flux through
the system.
We describe how our results can be
related to systems that exhibit jamming.  Since a self-jamming state
appears in the limit of high activity or sufficiently large obstacle density,
both activity and disorder strength act
as two additional parameters that can produce jamming in particle assemblies.

\acknowledgments
This work was carried out under the auspices of the 
NNSA of the U.S. DoE at LANL under Contract No.
DE-AC52-06NA25396.

\end{document}